\documentclass[a4paper,fleqn]{cas-sc}

\usepackage[authoryear,longnamesfirst]{natbib}

\def\tsc#1{\csdef{#1}{\textsc{\lowercase{#1}}\xspace}}
\tsc{WGM}
\tsc{QE}
\tsc{EP}
\tsc{PMS}
\tsc{BEC}
\tsc{DE}

\begin{document}
\let\WriteBookmarks\relax
\def\floatpagepagefraction{1}
\def\textpagefraction{.001}


\shortauthors{Behzadi et al}  

\title [mode = title]{Weakly-Supervised Deep Learning Model for Prostate Cancer Diagnosis and Gleason Grading of Histopathology Images}  



%

\author[1,3]{Mohammad Mahdi Behzadi}

\cormark[1]


\ead{mohammad.behzadi@uconn.edu}



\affiliation[1]{organization={Department of Computer Science, University of Connecticut},
            addressline={371 Fairfield Way, Unit 4155}, 
            city={Storrs},
            postcode={06269-4155}, 
            state={CT},
            country={USA}}

\author[1,3]{Mohammad Madani}

\cormark[1]
\ead{mohammad.madani@uconn.edu}




\author[2]{Hanzhang Wang}

\ead{hanzwang@uchc.edu}



\affiliation[2]{organization={Pathology and Laboratory Medicine, University of Connecticut Health Center},
            addressline={300 UConn Health Boulevard}, 
            city={Farmington},
            postcode={06030}, 
            state={CT},
            country={USA}}

\author[1]{Jun Bai}
\ead{jun.bai@uconn.edu}

\author[1]{Ankit Bhardwaj}
\ead{ankit.bhardwaj@uconn.edu}

\author[3,4]{Anna Tarakanova}
\ead{anna.tarakanova@uconn.edu}
\affiliation[3]{organization={Department of Mechanical Engineering, University of Connecticut},
            addressline={191 Auditorium Rd. U-3139}, 
            city={Storrs},
            postcode={06269}, 
            state={CT},
            country={USA}}

\affiliation[4]{organization={Department of Biomedical Engineering, University of Connecticut},
            addressline={263 Farmington Ave}, 
            city={Farmington},
            postcode={06030}, 
            state={CT},
            country={USA}}

\author[2]{Harold Yamase}
\ead{yamase@uchc.edu}

\author[2]{Ga Hie Nam}
\ead{gnam@uchc.edu}

\author[1]{Sheida Nabavi}

\cormark[2]
\ead{sheida.nabavi@uconn.edu}

\cortext[1]{Equal Contribution}
\cortext[2]{Corresponding author}


\begin{abstract}
Prostate cancer is the most common cancer in men worldwide and the second leading cause of cancer death in the United States.  One of the prognostic features in prostate cancer is the Gleason grading of histopathology images. The Gleason grade is assigned based on tumor architecture on Hematoxylin and Eosin (H\&E) stained whole slide images (WSI) by the pathologists. This process is time-consuming and has known interobserver variability. 
In the past few years, deep learning algorithms have been used to analyze histopathology images, delivering promising results for grading prostate cancer. However, most of the algorithms rely on the fully annotated datasets which are expensive to generate. 
In this work, we proposed a novel weakly-supervised algorithm to classify prostate cancer grades. The proposed algorithm consists of three steps: (1) extracting discriminative areas in a histopathology image by employing the Multiple Instance Learning (MIL) algorithm based on Transformers, (2) representing  the image by constructing a graph using the discriminative patches, and (3) classifying the image into its  Gleason grades by developing a Graph Convolutional Neural Network (GCN) based on the gated attention mechanism.
We evaluated our algorithm using publicly available datasets, including TCGA-PRAD, PANDA, and Gleason 2019 challenge datasets. We also cross validated the algorithm on an independent dataset. Results show that the proposed model achieved state-of-the-art performance in the Gleason grading task in terms of accuracy, F1 score, and cohen-kappa. The code is available at https://github.com/NabaviLab/Prostate-Cancer.
\end{abstract}



\maketitle

\section{Introduction} \label{introduction}
Prostate cancer is the most common type of cancer in men and the second leading cause of cancer death in the United States. \cite{siegel2021cancer}. Accurate pathological diagnosis and Gleason grading of prostate cancer are very important in prognostication and treatment decisions by the urologists. \cite{epstein20162014} The Gleason grade is assigned based on tumor architecture on Hematoxylin and Eosin (H\&E) stained whole slide images (WSIs) by the pathologists. This process is time-consuming and has known interobserver variability. \cite{allsbrook2001interobserver}. Recently, many Computer-Aided Diagnosis (CAD) algorithms have been developed to help physicians make the correct decision and reduce human error in cancer diagnosis. However, the output of CAD systems needs to be evaluated by physicians because these systems usually detect more false features than true marks. This evaluation increases the reading time and limits the number of cases that pathologists can evaluate \cite{yanase2019seven, madani2022role}. Recently, the advancement of deep learning (DL) algorithms along with the availability of large medical datasets, has effectively sped up the image analysis process and helped pathologists in early cancer diagnoses, even in some cases, the DL algorithms outperformed human pathologists \cite{bejnordi2017diagnostic}.

Although DL methods show promising results in cancer diagnosis and grading, their thirst for labeled data prevents them from being used in practice since labeling the data should be done by clinical experts and it is very time-consuming. The histopathology images used for cancer diagnosis, are very large images (e.g, in order of gigapixel) that cannot be directly used by DL algorithms. To make them suitable for DL methods, the images need to be cropped into small patches. Each WSI can have hundreds of patches and each patch should be annotated in order to be used in DL algorithms. Moreover, each patch in an image can have a different label since some parts of the image can be benign or normal, and other parts can be cancerous areas. An expert pathologist is needed for annotating the patches in WSIs. This can be very time-consuming and expensive. 
 To address these limitations, weakly supervised algorithms based on multiple-instance learning (MIL) have been developed for cancer diagnosis using histopathology images. Generally, in these algorithms, a WSI is divided into small patches creating a bag of all the patches. Each bag has an assigned label (i.e., the label of the WSI), but the model does not have access to the label of each patch in the bag. The MIL approach has been successfully applied to WSI-level cancer detection \cite{campanella2019clinical} and Gleason grading of WSIs \cite{silva2021self}. Although these methods showed promising results, they include the non-discriminative patches in the classifier which may decrease the performance. Moreover,  they do not take into account the relation between the cancerous patches in WSIs. 
  
 To address these limitations, in this work we proposed a weakly supervised algorithm to perform the Gleason grading classification in histopathology images. We developed an MIL algorithm based on the Transformers \cite{vaswani2017attention} to assign a score to each patch in the bag based on its effect on the label of the bags. To make the training more efficient, instead of training on thousands of patches, we extracted the features of each patch using an autoencoder that was trained to reconstruct the patches.
 The model is trained on the bag of feature vectors obtained by feeding an autoencoder model with the patches in the bag. Then, we used the top-scored patches to construct a graph whose nodes are the patches and the edges are obtained by connecting the patches to their nearest neighbors based on the location of each patch in the original slide. The constructed graphs along with the feature vectors corresponding to their node were fed to a Gated-Attention Graph Neural Network for Gleason grade classification. We evaluated the performance of the proposed method in terms of accuracy, F1 score, and cohen-kappa using the TCGA-PRAD, PANDA, and Gleason 2019 challenge dataset. Results show that our model outperforms the state-of-the-art baseline models by significant margins.    Key contributions of our proposed model can be listed as: 1) deriving feature maps for each patch of WSIs by using an autoencoder to increase the efficiency of the model training;
 2) extracting  discriminative patches within  WSIs via a Transformer-based network architecture;
 3) representing each WSI by a graph whose nodes are discriminative patches within the WSI;
 4) developing a graph neural network model called augmented self-attention graph neural network (ASG) to obtain the Gleason grade of WSIs;
 5)obtaining the best performance compared to the state-of-the-art models using three different datasets. 

\section{Related Works} \label{relatedworks}
The Gleason scoring system, introduced in 1966, is a grading system for prostatic carcinoma based solely on the architectural pattern of the tumor. The grading system has undergone multiple modifications and the current modified Gleason grading encompasses patterns 3 to 5   \cite{gleason1992histologic}. Pattern 3 consists of discrete glandular units with well-formed lamina whereas pattern 4 consists of poorly formed, cribriform, fused, and hypernephromatoid glands. Pattern 5 represents the most poorly differentiated tumor with no glandular differentiation composed of solid sheets, cords, single cells, or with comedonecrosis.  The final Gleason score is the sum of the majority (primary pattern) and minority (secondary) growth patterns. The Gleason grade group (GGG) would be 1 (GGG1) if the sum $\leq$6, GGG2 (3+4=7)and GGG3 (4+3=7) if the sum is 7, GGG4 if the sum is 8 (4+4, 3+5, 5+3) and GGG5 if the sum is 9-10 (4+5, 5+4, 5+5) \cite{epstein20162014}.  

Finding the Gleason score by pathologists can be challenging since it highly depends on the experience of the expert, visual perception and cognitive ability \cite{fitzgerald2001error}. Moreover, some of the Gleason patterns are hard to differentiate i.e.  GGGs of 3 and 4 \cite{latour2008grading}. Some algorithms have been proposed to extract additional features from the tissue and help physicians to find the grade more accurately. For example, the proposed method in \cite{smith1999similarity} used Fourier transform to extract the texture components of the images. Then, the nearest neighbor classifier was used to find the grade of each image. In another work \cite{tabesh2007multifeature}, the color, texture, and morphometrical features of the images at the global and histological object levels were extracted and combined for classification. The authors compared the performance of Gaussian, nearest neighbor, and support vector machine classifiers together with the sequential forward feature selection algorithm. Their method classifies the images into low and high-grade classes with an accuracy of 81.0\%. In another work \cite{jafari2003multiwavelet}, the energy and entropy features of multiwavelet coefficients of the image were used along with a k-nearest neighbor classifier to classify each image into grades 2 through 5. The authors in \cite{nguyen2012prostate} extracted tissue structural features from gland morphology. Using the extracted features, they could classify a tissue pattern into three major categories: benign, grade 3 carcinoma, and grade 4 carcinoma with 85.6\% accuracy. Despite having good results, these methods heavily depend on accurately identifying   small  patches or ROIs from which effective features are extracted. This is a drawback  since identifying ROIs is expensive and  requires domain experts. Moreover, the good results reported in these studies are due to their reliance on good feature extraction and these studies suffer from subjectivity and limited performance on the independent dataset. In this situation,  deep learning methods can be employed to reduce errors and increase generalizability.

Deep learning algorithms have shown an outstanding capability to discover complex patterns in different types of data, either sequential data \cite{kunkel2021modeling, madani2021dsressol,madani2022improved} or images \cite{behzadi2021real, behzadi2022gantl,liu2022machine} including medical images \cite{li2018path, bai2021applying,bai2022feature,bai2022applying,bai2022semi}. This capability reduces the need for using handcrafted features by extracting discriminative features from images. For example, Wenyuan et al. \cite{li2018path} proposed a region-based convolutional neural network (CNN) framework where they extracted feature maps for each patch using ResNet \cite{he2016deep}, then they developed a two-stage training strategy to detect epithelial cells and predict Gleason grades simultaneously using the feature maps. In other work done by \cite{arvaniti2018automated}, the authors developed a CNN model based on the MobileNet \cite{howard2017mobilenets} architecture as a patch-level classifier. Then, the trained model was applied to entire images in a sliding window fashion and generated pixel-level probability maps for each class. The fully annotated dataset of prostate cancer tissue microarrays (TMA) \cite{zhong2017curated} has been used for this model. Lucas et al. \cite{lucas2019deep} developed a CNN model based on the Inception v3 \cite{szegedy2016rethinking} architecture using small patches extracted from a well-annotated dataset to distinguish Gleason Grade 3 and Gleason Grade higher than or equivalent to 4 from non-atypical various (including both healthy glands and glands with low-grade prostatic intraepithelial neoplasia) prostate histopathology tissue. Their model could differentiate non-atypical from malignant (Gleason grade $\geq 3$) areas with an accuracy of 92\%, and Gleason grade$ \leq 4$ from Gleason grade $\leq 3$ with an accuracy of 90\%.
In another work proposed by Gour et al. \cite{gour2022application}, the authors compared the performance of the state-of-the-art CNN models for prostate cancer grading using histopathological images. They resized  WSIs to $512 \times 512$ patches and then cropped them according to the network input sizes. According to their results, the EfficientNet \cite{tan2019efficientnet} obtained the best performance compared to the other CNN architectures.
Although these approaches provided a reasonable accuracy, they heavily relied on pixel-level annotated datasets which should be done by experts and are very time-consuming. To reduce the need for annotating the data, Otalora et al. \cite{otalora2021combining} proposed a transfer learning approach in which a CNN model is pretrained on the patches extracted from a fully annotated (region annotations) dataset and is later fine-tuned on a dataset with less expensive weak (image-level) labels. Their results showed that the model performance increases when fine-tuning a pretrained model on a fully annotated dataset for the task of Gleason scoring with the weak WSI labels. Despite of a good performance, the framework proposed in \cite{otalora2021combining} still needs a large fully annotated dataset for their pretrained model. To address this limitation, weakly supervised algorithms have been developed for the Gleason grading of prostate cancer.

The weakly supervised algorithms used for Gleason grading can be divided into two categories: 1) some papers developed a semi-supervised algorithm to annotate WSIs, and 2) other papers used MIL to train their model. For example, a semi-supervised technique based on Active Learning was developed by Singhal et al. \cite{singhal2022deep} to produce pixel-level annotations of WSIs. In their approach, first, a pathologist selected and annotated a limited fraction of images, then, a CNN model was trained to segment the tumor locations. In the next step, the system simulated labels using iterative active learning  (inspired by the Cost-Effective Active Learning paradigm \cite{gorriz2017active})  to assign a label to the patches. Finally, the authors iteratively trained a U-Net-based \cite{ronneberger2015u} CNN for Gleason grade group identification in which in each iteration, the unlabeled patches are subsequently fed into the trained CNN, and a measure of uncertainty was computed for each unlabeled patch. A pathologist then annotated the most uncertain samples (with uncertainty measures more than a threshold) and adds them to the training set. In the following iteration, the CNN was re-trained with a new set of annotated images. Despite having a good performance, their approach is very time-consuming since it has to train the model over and over until the maximum uncertainty is less than the threshold. Moreover, it still needs the help of a pathologist in each iteration. Furthermore, their algorithm may have some miss classifications of patches in the WSI which may affect the overall performance of the deep learning classifier. The methods that use MIL do not need region-level annotations. For example, in a work done by Campanella et al. \cite{campanella2019clinical}, the authors first cropped the tissue images into small tiles, then they developed an MIL training procedure to rank the tiles according to their probability of being positive. Finally, the top-ranked tiles in each slide were sequentially passed to the recurrent neural network (RNN) to predict the final slide-level classification. Although their method showed a promising result, they do not take into account the information aggregation from the cancerous patches in WSI.   
  
 In this work, we propose a weakly supervised algorithm that first extracts  discriminative patches in  WSIs, then by representing the cancerous patches as a graph, employs a graph convolutional neural network (GCN) approach to aggregate neighbor patches' information for classifying Gleason grades. 
 We show that constructing a graph using  cancerous patches and employing GCN increases the accuracy of grading classification.

\section{Methods}
The proposed framework contains three main steps. The first step is preprocessing  WSIs and cropping them into small patches. The second step is  training an autoencoder using the obtained patches, and using the trained autoencoder to extract the feature vector for each patch. The final step is  training the classifier using the feature vectors to classify the grades. The classifier is made of three parts. The first part assigns  scores to all feature vectors based on their importance in the labels of the WSIs. The second part creates a graph based on the top-scored feature vectors of each WSI using the nearest neighbor algorithm. In the last step, a GCN is trained to classify the graph to a proper Gleason grade group. Figure \ref{fig:overview} shows the overview of the proposed framework.

\subsection{Data preprocessing}

Applying a CNN directly on WSIs is computationally expensive and in some cases impossible due to the size of WSIs. We used the idea of MIL to overcome this problem in which each WSI is a collection (bag) of instances (small patches). The detail of the model and training process is described in section \ref{model development}. To construct the bags, we divided each WSI into $M$ small patches of size $ N \times N$. 
Since a large area of  WSIs is the background pixels, we filtered out background patches by imposing a criterion to select patches  that at least $z\%$ of their pixels are from the tissue pixels. 
Training the model on thousands of patches is very time-consuming. To accelerate the training process, we reduce the dimension of each bag by feeding each patch to  an autoencoder and used their latent vectors of size $2560$ to represent patches.
Accordingly, each WSI  is represented by a $M \times 2560$ matrix.
\subsection{Feature extraction}
To extract the feature vectors of the instances in each bag, we implemented an unsupervised autoencoder model. The encoder part is a pre-trained EfficientNetB7 model \cite{tan2019efficientnet} that has shown superior performance in Gleason grading compared to other pre-trained models \cite{gour2022application}. The decoder part contains six building blocks each consisting of a convolutional layer and a convolutional transpose layer. To train the model, we use the mean squared error between the reconstructed input and the original input as the loss function. 
The extracted features at the encoders are used to represent patches.
These features are used to identify discriminative patches 
in the first stage of the proposed model and the node features  in the second stage of the proposed model.

\subsection{Classification model} \label{model development}

The label of the patches in a WSI can be different from each other and from the label of the WSI. 
Owing to the lack of labeled patches for each WSI, we cannot feed the patches directly to a CNN classifier model. To deal with this problem, we borrow the idea of MIL \cite{lu2021data}. To build our MIL-based architecture, we utilize a two stages model created via the Transformer and GCN models. In the first stage, we try to capture key discriminative patches (instances) within each WSI (bag); then, we represent each WSI (bag) as a graph, the nodes of which are the discriminative patches. Finally, in the second stage of the model, we feed the constructed graphs to a GCN model to classify Gleason grades. 

\subsubsection{Extracting discriminative patches}
As mentioned above, we extract patches with the size of $N \times N$ for each WSI. Some patches contain discriminative information to determine the label of a WSI and some of them may contain less information. To find the most informative patches, we adopt a novel Transformer-based model. Our Transformer includes a projection ($f_{p}$) and an attention ($f_{at}$) modules. 

The projection module contains a series of trainable fully-connected layers projecting the fixed feature embeddings derived from the latent space of our proposed autoencoder model into a more compact feature space. Thus, given the bag of $H_{i}$ that contains $M_i$ feature vectors of size $1 \times 2560$ extracted from  the $i^{th}$ WSI, 
the projection layer $W_p \in R^{512 \times 2560}$ is employed to project patch embeddings to a 512-dimensional feature space. As an important note, each row ($m_i$) of  $M_i$ is related to a single patch of the WSI. This feature space is fed to an attention module ($f_{at}$) consisting of a series of self-attention pooling mechanisms inspired by the Transformer-based architecture introduced by Ilse et al \cite{ilse2018attention}. The attention module tries to capture information-rich patches within a WSI, which are key discriminative feature maps for the prediction of the Gleason score of WSI. In other words, our main objective to use the attention module is to rank the most discriminative patches based on their attention scores. Attention module includes 3 fully-connected layers with initial weights of $K_a\in R^{1 \times 256} $, $Q_a\in R^{256 \times 512}$, and $V_a\in R^{256 \times 512}$  to calculate the contribution of each patch in final label of WSI. These weights learn to assign attention score ($a_{i,h}$) for $h^{th}$ patch in the $i^{th}$ WSI based on the incoming patch embedding $m_{i,h}$ via following equation:
\begin{equation}
    a_{i,h} = \frac{exp\{ K_{a} (tanh(V_{a}m_{i,h}^{T}) \cdot sigm(Q_{a}m_{i,h}^{T}))\}   }{\sum_{h=1}^{H_i} exp\{ K_{a} (tanh(V_{a}m_{i,h}^{T}) \cdot sigm(Q_{a}m_{i,h}^{T}))\}  },
\end{equation}
where $tanh$ and $sigm$ are tangent hyperbolic and sigmoid activation functions, respectively.
Thus, the final attention score vector for a WSI is denoted by $A_i=f_{at}(M_i)$. 
In existing Transformer-based methods \cite{lu2022federated}, after calculating the attention score vector, the bag-level representation $A_{bag,i}$ is calculated by utilizing  the predicted attention scores as weights for averaging all the feature embeddings in the bag as $A_{bag,i}=\ \sum_{h=1}^{H_i}a_{i,h}m_{i,h}$. Then,  $A_{bag, i}$ is fed to a simple fully connected layer (softmax activation) to predict the slide-level label. 
This approach (we used it as one of our baseline models) can generate erroneous inference,  because it considers all patches regardless of their scores which can add  noise  to the prediction process. Thus, to deal with this issue and improve our model’s performance, patches with low attention scores are pruned and top-score patches are selected for the second stage of the model.

\subsubsection{Graph generation}
We selected the top $S\%$  attention scored patches for each WSI as highly discriminative patches. To effectively aggregate the discriminative patches information we represent WSIs as graphs and employ GCN to classify each graph into the appropriate label.
To build the graph, each discriminative patch is considered as a node for the graph, and each node is connected to its $K$ nearest neighbors. The node features (patch feature vectors) along with the edge information were used in the second stage of the model.

\subsubsection{Augmented Self-Attention Graph Convolutional Neural Network (ASG) model}
\begin{figure*}[th!]
    \centering
    \includegraphics[width=\textwidth]{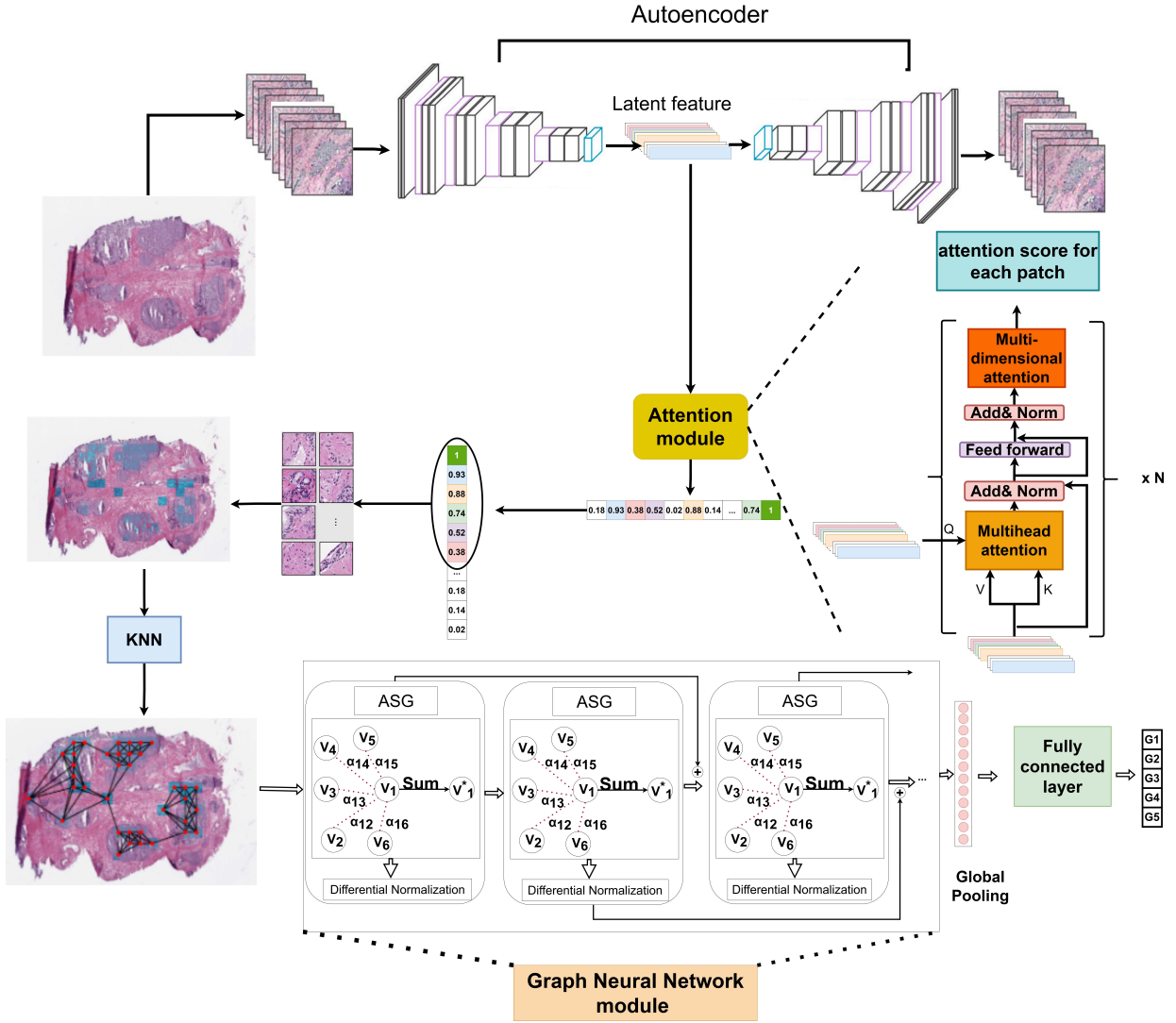}
    \caption{The overview of the proposed algorithm. A WSI is first cropped in small patches, then the feature vector for each patch is extracted using the  autoencoder. The extracted feature vectors are fed to the first stage of the model to predict the score corresponding to each patch. Using the high-score patches, we construct a graph using KNN. Finally, in the second stage of the model, a GCN is used to classify the graph into a Gleason grade group.}
    \label{fig:overview}
\end{figure*}

The main idea of our model is to better represent a node feature by aggregating features of its neighboring nodes and using the self-attention mechanism to reinforce each node to focus on itself. Our model includes 8 augmented self-attention graph neural network (ASG) modules \cite{lee2019self} followed by an average pooling and a fully connected layer (Figure \ref{fig:overview}). The ASG modules contain one single graph convolutional network (GCN) layer and one self-attention mechanism. For each GCN layer within the ASG module, the propagation process recursively aggregates neighboring nodes' features to derive the new nodes' feature map via the following equation \cite{kipf2016semi}:
\begin{equation}
    G^l=ReLU(D^{-\frac{1}{2}}(A^{l-1}+I)D^{-\frac{1}{2}}G^{l-1}W_G^{l-1}),
\end{equation}

where $A^{l-1}$ is the adjacency matrix of layer $(l-1)$, D is a diagonal degree, $W_G^{l-1}$ is a matrix of trainable weights in layer $(l-1)$, and $G^{l-1}$ is the node feature aggregated matrix at layer $(l-1)$.  Also, the self-attention mechanism within the ASG module tries to select top $N^l=d \times N^{l-1}$ nodes, in which $d\in (0,1)$. Thus, attention scores are obtained as follows \cite{kipf2016semi}:
\begin{equation}
    T^l=ReLU(D^{-\frac{1}{2}}(A^l+I)D^{-\frac{1}{2}}G^lW_T^l),
\end{equation}

where $W_T^l$ is a learnable weight matrix. When attention scores are calculated, the new node features ($X_V^l$) and adjacency matrix ($A^l$) for top nodes ($N^l$) are computed via the following equations \cite{cangea2018towards}:
\begin{equation}
    X_V^l=\left\{X_1^l,X_2^l,\ldots,\ X_{N^l}^l\right\}=index\ (X_V^{l-1})\cdot T_{idx}^l,
\end{equation}
\begin{equation}
    A^l\in R^{N^l\times N^l}=index(A^{l-1}),
\end{equation}

where $Index(X_V^{l-1})$ returns node-wise indexed feature matrix, $T_{idx}^l$ is feature attention scores matrix, and  $A^l$ returns row and column-wise indexed adjacency matrix. Our ASG can extract comprehensive feature maps from the input graph and provide interpretable results in each node’s contribution. With the expectation that a deeper model could extract even more of these inter-node-dependent feature maps, we devised our model, such that it  uses additive skip connections between the ASG modules to extract  more discriminative local features. In the next step, in order to flatten the graph feature matrix to the same size feature for the classification process, we utilize the average pooling layer. Finally, by passing this flattened feature vector through the fully-connected layer, we predict the slide-level label of each WSI. For the training process, we used categorical cross-entropy loss as shown below: 
\begin{equation}
    l=\ -\sum_{i=1}^{n}t_i\log (p_i),
\end{equation}
 
where $n$ is the number of classes, $t_i$ is the truth label and $p_i$ is the softmax probability for the $i^{th}$ class.
Figure \ref{fig:overview} shows the framework of the proposed method.

\section{Datasets and Experiment}
\subsection{Datasets}
We use four datasets to evaluate the performance of the proposed model as described below.\\
\textbf{The GLEASON 2019 challenge dataset} \cite{nir2018automatic, karimi2019deep}. This dataset is part of the grand challenge for pathology at the International Conference on Medical Image Computing and Computer-Assisted Intervention (MICCAI) 2019. The dataset contains 244 prostate TMA images for training and 88 TMA images for testing purposes. Six pathologists with different levels of experience were asked to annotate the dataset at the pixel level. Four pathologists annotated all 333 TMA images but the other two only annotated 191 and 91 of the images, respectively. The final labels were obtained by pixel-wise majority voting. We only used the 244 training images for training and testing our model because the ground truth labels were not available for the remaining test images.

\textbf{TCGA-PRAD.} This dataset was obtained from the  National Cancer Institution. The dataset contains 723 cancer tissue slides from 490 patients and does not provide benign prostate slides. 
The patients' clinical information contains the primary and secondary Gleason pattern for each patient and the final GGG was obtained by summing the primary and secondary scores. 

\textbf{Prostate Cancer Grade Assessment (PANDA) Challenge.} The PANDA dataset is the largest available public dataset \cite{bulten2020panda} containing 10,616 biopsy images divided into six classes. The biopsy images were collected by Karolinska Institute and Radboud University Medical Center with the collaboration of several pathologists. The dataset contains the primary and secondary grades along with the GGG of each slide.

\textbf{University of Connecticut Health Center (UCHC) dataset.} UConn Health Pathology Prostate Cancer database collected 30 prostate cancer WSI from 16 de-identified radical prostatectomy cases over May 2018 to September 2022. Two board certificated  pathologists with Genitourinary specialty interest (GN and HY) independently provided primary, secondary Gleason scores and Gleason Group on each WSI. All prostatectomy cases were performed for clinical purposes at UConn Health John Dempsey Hospital (IRB number 23-108-1). This dataset is used for cross validation only.

\subsection{Baseline models} \label{baselines}
We compared the proposed model with the following
baseline models.\\
\textbf{Proposed model + No AE:} In this model, we used pretrained EfficientNet trained on the ImageNet dataset (without any additional training) instead of the autoencoder to extract the feature vector for each instance. Everything else is the same as the proposed model.

\textbf{Proposed model\_att:} In this model, the output of the first stage of the model is directly fed to a fully connected layer instead of using the second stage of the proposed model, to classify the grades.

\textbf{\cite{otalora2021combining}:} The model proposed in \cite{otalora2021combining} was trained on the Gleason 2019 dataset, and their weakly supervised model was used for the TCGA-PRAD dataset. As described in Section \ref{relatedworks}, the model predicts the primary and secondary scores of the patches in the slide, and WSI labels  are computed by taking the majority voting of the most frequently predicted Gleason patterns. We used the code provided by the authors with the default training setting.

\textbf{\cite{arvaniti2018automated}:} The model is based on the MobileNet architecture. We trained the given model on the Gleason 2019 dataset for patch-wise grading. The slide labels were generated based on the primary and secondary scores predicted by the model. We used the code provided by the authors with the default training setting.

\textbf{EfficientNet-B7 , ResNet50,} and \textbf{InceptionV3:} We followed the approach presented in \cite{gour2022application} to fine-tune these classifiers for our grade classification task.  These models have been pretrained on the ImageNet dataset. To fine-tune the models, the top layer has been replaced with a custom classifier.

\section{Results and Discussion}
\subsection{Overall results}
\begin{figure*}[th!]
    \centering
    \includegraphics[width=\textwidth]{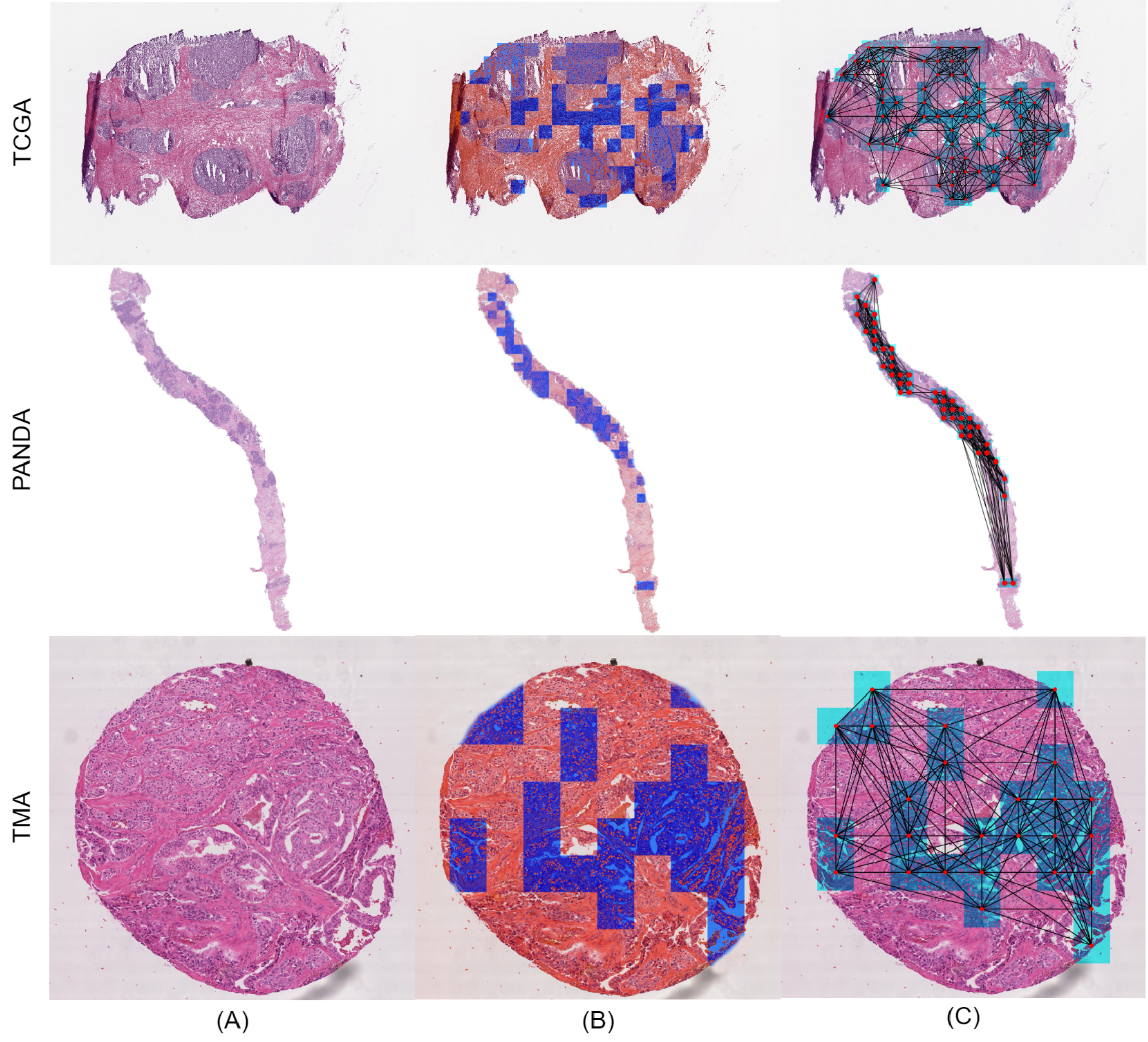}
    \caption{The discriminative patches extracted by the first stage of the model for three datasets. Column (A) is the original slides, column (B) shows the discriminative patches as blue rectangles in the slides, and column (C) depicts the  constructed graphs using  the discriminative patches as nodes.}
    \label{fig:graphs}
\end{figure*}

We first evaluated the selected discriminative patches by manually examining the slides and patches, conducted by two pathologists in our team.
We observed that the discriminative areas  extracted by the model are very relevant to the cancerous areas in the slides. Figure \ref{fig:graphs} shows an example of the original slide, the discriminative patches, and the graph constructed on top of the extracted patches for the three datasets. Column (A) in Figure \ref{fig:graphs} shows the original image, the discriminative patches are shown with dark blue in column (B), and the constructed graphs are shown in column (C). The expert pathologists confirm that a majority of the selected high attention patches fall into WSI region of interest (ROIs). Since the selected patches do not cover all the ROIs of a WSI 
we called them  discriminative regions, not ROIs.    Note that we select top $S\%$ high attended patches and the number of discriminative patches changes with changing this threshold to cover ROIs. However, for classifying WSIs we do not need all the ROIs and the most discriminative patches can be used for classification without adding noise to the model.

\begin{figure*}[th!]
    \centering
    \includegraphics[width=\textwidth]{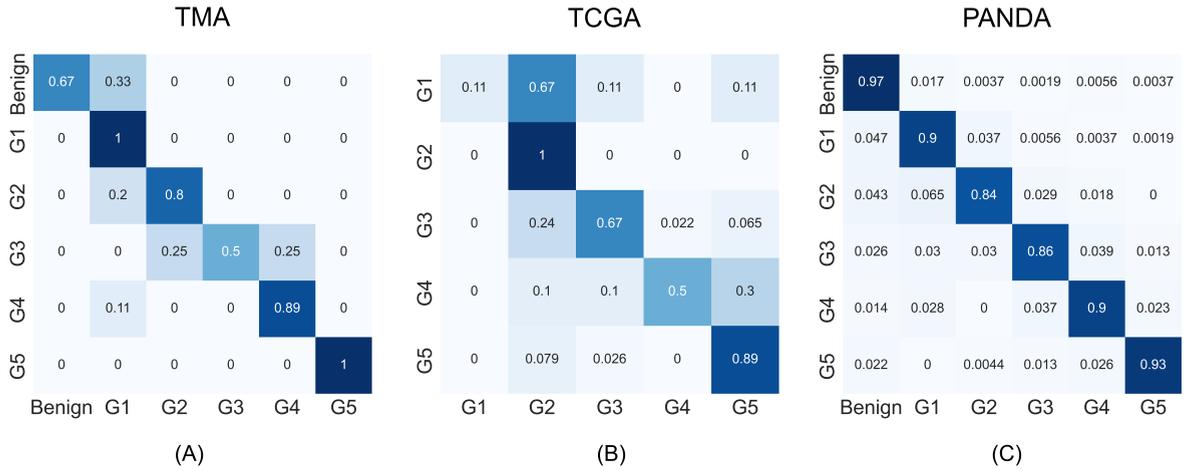}
    \caption{The confusion matrix for all the datasets.}
    \label{fig:confusion}
\end{figure*}

We also compared the performance of the proposed model to the baseline models explained in section \ref{baselines} in terms of accuracy, F1-score, sensitivity, precision, and Cohen-kappa on all three datasets.
Table \ref{tab:TMA} shows the performance comparison between the proposed method and the baseline models for the Gleason 2019 challenge dataset. We used Five-fold cross-validation to train the model. The proposed method  outperforms the baseline models in accuracy, F1-score, sensitivity, and cohen-kappa. Moreover, it can be seen that using the autoencoder to extract patch features can improve the accuracy by about $1\%$. One of the main challenges of the Gleason 2019 challenge dataset that can affect the performance of the model is that the dataset contains only 244 data and the model can get overfitted easily. Moreover, since the training was done on the 5-fold cross-validation, the number of test data for this dataset was much lower than the others. Furthermore, the dataset is highly imbalanced which may make predicting  correct labels for certain groups difficult. For example, as shown in the first column of Figure \ref{fig:confusion} (Figure \ref{fig:confusion}.A), the model has difficulty predicting labels of the first and fourth groups, which contain less number of samples. However, it could predict all the data in the second group correctly. 

We  repeated the same process for the other two datasets as well. Table \ref{tab:TCGA} shows the result of comparing the proposed model and the baselines for the TCGA-PRAD dataset. As shown in Table \ref{tab:TCGA}, the proposed method outperforms the baseline models in most of the evaluation metrics. The TCGA-PRAD dataset is a larger dataset compared to the Gleason 2019 challenge dataset and the model is less likely to be overfitted. Moreover, due to the size of the slides, each bag in the TCGA-PRAD dataset contains more instances than the other two datasets. Therefore, the number of discriminative patches would be higher, resulting in   more complex graphs compared to those of   the other two datasets. However, like the Gleason 2019 challenge dataset, this dataset is imbalanced and the model has difficulty predicting the labels of the groups  including a lower number of data. As shown in  Figure \ref{fig:confusion}.B, the model can predict the label of the second group with $100 \%$ accuracy, but it doesn't perform well in classifying the first group. 

For the PANDA dataset we performed our experiment for both the Gleason group classification and the binary classification since there are relatively large numbers of benign and malignant in the dataset. The results of comparing the proposed method with the baseline models for grade classification and the binary classification are shown in Table \ref{tab:PANDA} and Table \ref{tab:PANDAbin}, respectively. The results indicate that our method outperforms the baseline models in both grade classification and binary classification. The PANDA dataset is a very large dataset, therefore, the model had a chance to learn all the labels relatively well. Figure \ref{fig:confusion}.C shows the confusion matrix for the PANDA dataset. It can be seen from the confusion matrix that the model performs relatively well in all labels.

\begin{figure*}[th!]
    \centering
    \includegraphics[width=\textwidth]{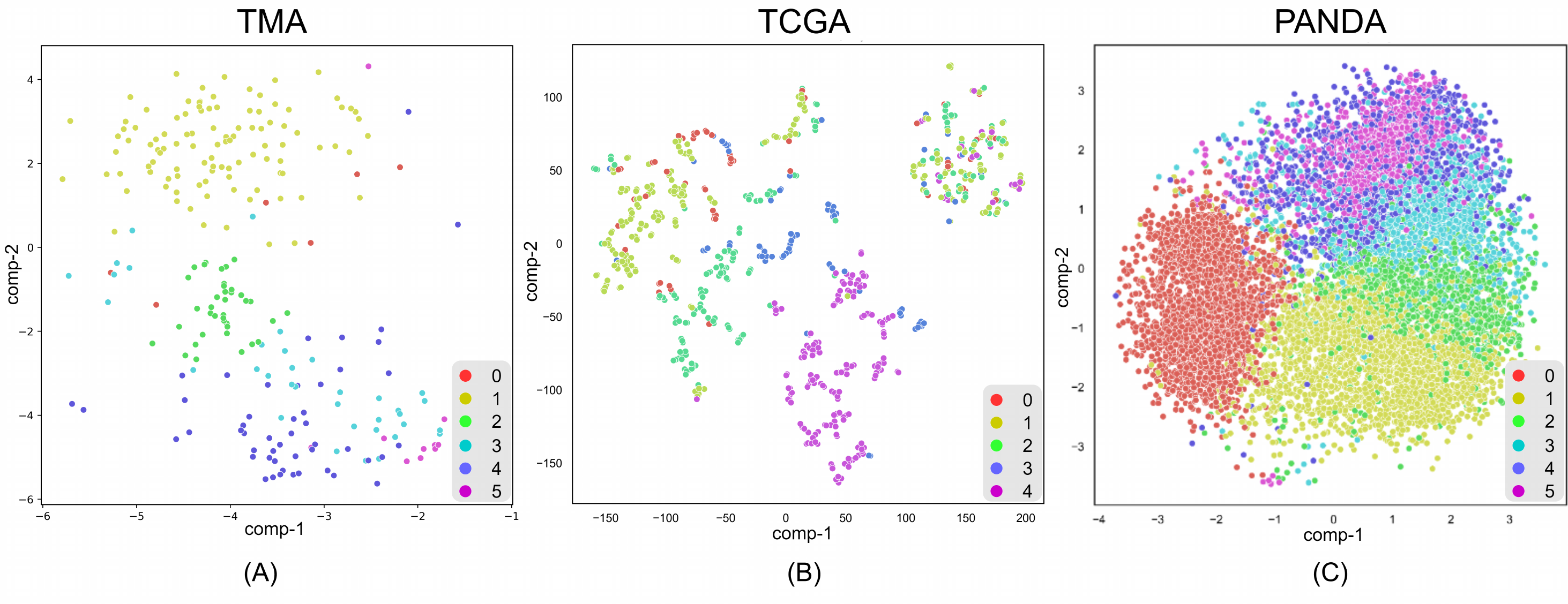}
    \caption{The TSNE on the output of the last layer before the final classifier layer for all the datasets.}
    \label{fig:tsne}
\end{figure*}

To show that the proposed model is trained effectively, we applied TSNE on the output of the last layer before the classifier layer. Figure \ref{fig:tsne} shows the result of TSNE for all the datasets. For the TCGA-PRAD dataset, the model can differentiate groups 4, 2, and 1 from each other, but it has difficulties separating groups 3 from  groups 4 and 2, and group 0 from group 1. In the Gleason 2019 dataset, the model can easily differentiate groups 2, 4, and 1 from each other. However, differentiating groups 3 and 4 from others seems to be difficult for the model. Finally, for the PANDA dataset, it seems that distinguishing groups 4 and 5 from each other is difficult for the model. However, the model can differentiate the other labels from each other with high confidence.   

To further evaluate the performance of our model, we test the model on our internal dataset, i.e., UCHC dataset as cross validation. The essence of this data is very close to TCGA dataset, so we only used the model trained on TCGA dataset to classify the grade of UCHC images. The dataset includes 15 slides in group 5, an average of 5 slides in each group 2 to 4 and no slide for group 1. The proposed model provides overall (considering all the groups) accuracy of 60\%, F1 score of 59.66\%, sensitivity of 100\%,  precision of 72.93\%, and Cohen-kappa of 0.37. As expected our model performs very well on label 5 with an accuracy of 80\%. Note that this group is also the only group with reasonable number of samples for having accountable results. The model does not perform well for group 2,  despite its good performance on TCGA dataset group 2 data (Figure \ref{fig:confusion}.B). This can be because we only have 6 samples in this group and the results cannot be statistically reliable. This can be true for labels 3 and 4, resulting in the overall lower performance compared to those for the TCGA dataset.  

\begin{table*}[ht!]
    \centering
    \caption{Comparison between the proposed model and the baseline model on the grade classification of Gleason 2019 challenge dataset}
    \resizebox{\columnwidth}{!}{
    \begin{tabular}{lcccccc}
        \toprule
        Metrics    & \shortstack{Proposed \\+  AE} & {\shortstack{Proposed \\+ No AE}} & {Proposed\_att}  & {\shortstack{\cite{arvaniti2018automated}}} & {\shortstack{ResNet50}} & {\shortstack{\cite{otalora2021combining}} }\\
        \midrule
        {\shortstack{Accuracy}}  & \textbf{0.845} & 0.829 & 0.764 & 0.794 & 0.729 & 0.817     \\
        F1-score  & \textbf{0.849}  & 0.794  & 0.733 & 0.769 & 0.733 & 0.795  \\
        Sensitivity   & \textbf{0.96} & 0.799  & 0.83  & 0.88 & 0.82 & 0.9  \\        
        Precision  & 0.764 & \textbf{0.773}  & 0.71  & 0.733 & 0.709 & 0.759    \\
        Cohen-kappa  & \textbf{0.797} & 0.755  & 0.58  & 0.733 & 0.593 & 0.767  \\
        
  		\bottomrule
    \end{tabular}
    }
    
    \label{tab:TMA}
\end{table*}

\begin{table*}[ht!]
    \centering
    \caption{Comparison between the proposed model and the baseline model on the grade classification of TCGA-PRAD dataset}
    \resizebox{\columnwidth}{!}{%
    \begin{tabular}{lccccccc}
        \toprule
        Metrics    & {\shortstack{Proposed \\ + AE}} & {\shortstack{Proposed \\+ No AE}} & {Proposed\_att}  & {\shortstack{\cite{otalora2021combining}}} & {\shortstack{ResNet50}} & {\shortstack{InceptionV3}}& {\shortstack{EfficientNet-B7}}\\
        \midrule
        {\shortstack{Accuracy}}  & \textbf{0.776} & 0.768 & 0.688 & 0.742 & 0.642 & 0.721 & 0.731     \\
        F1-score  & \textbf{0.753}  & 0.744  & 0.654 & 0.738 & 0.691 & 0.7 & 0.732  \\
        Sensitivity   & 0.66 & 0.68  & 0.582  & 0.651 & \textbf{0.732} & 0.63 & 0.679  \\
        Precision  & \textbf{0.832} & 0.783  & 0.794  & 0.802 & 0.652 & 0.764 & 0.794    \\
        Cohen-kappa  & \textbf{0.685} & 0.675  & 0.54  & 0.643 & 0.489 & 0.62 & 0.639  \\
        
  		\bottomrule
    \end{tabular}
    }
    \label{tab:TCGA}
\end{table*}

\begin{table*}[ht!]
    \centering
    \caption{Comparison between the proposed model and the baseline model on the grade classification of PANDA dataset}
    \resizebox{0.80\columnwidth}{!}{%
    \begin{tabular}{lcccccc}
        \toprule
        Metrics    & {\shortstack{Proposed \\ + AE}} & {\shortstack{Proposed \\ + No AE}} & {Proposed\_att}  & {\shortstack{EfficientNet-B7}} & {\shortstack{ResNet50}} & {\shortstack{InceptionV3}}\\
        \midrule
        {\shortstack{Accuracy}}  & \textbf{0.911} & 0.896 & 0.811 & 0.908 & 0.783 & 0.883     \\
        F1-score  & \textbf{0.91}  & 0.907  & 0.795 & 0.903 & 0.798 & 0.881  \\
        Sensitivity   & \textbf{0.953} & 0.928  & 0.832  & 0.938 & 0.82 & 0.9  \\
        Precision  & 0.909 & 0.874  & 0.779  & \textbf{0.912} & 0.766 & 0.874    \\
        Cohen-kappa  & 0.889 & 0.867  & 0.769  & \textbf{0.892} & 0.722 & 0.867  \\
        
  		\bottomrule
    \end{tabular}
    }
    
    \label{tab:PANDA}
\end{table*}

\begin{table*}[ht!]
    \centering
    \caption{Comparison between the proposed model and the baseline models on the binary classification (Benign Vs Malignant) of PANDA dataset}
    \resizebox{0.70\columnwidth}{!}{%
    \begin{tabular}{lccccc}
        \toprule
        Metrics    & {\shortstack{Proposed \\ + AE}} & {\shortstack{Proposed \\+ No AE}} & {Proposed\_att}  & {\shortstack{EfficientNet-B7}} & {\shortstack{ResNet50}} \\
        \midrule
        {\shortstack{Accuracy}}  & \textbf{0.957} & 0.949 & 0.876 & 0.948 & 0.895     \\
        F1-score  & \textbf{0.933}  & 0.928  & 0.846 & 0.93 & 0.862  \\
        Sensitivity   & 0.925 & 0.921  & 0.835  & \textbf{0.932} & 0.848  \\
        Precision  & \textbf{0.941} & 0.936  & 0.857  & 0.93 & 0.877    \\
        Cohen-kappa  & \textbf{0.885} & 0.865  & 0.785  & 0.859 & 0.811  \\
        
  		\bottomrule
    \end{tabular}}
    \label{tab:PANDAbin}
\end{table*}

\subsection{Ablation study}

To  examine the sensitivity and robustness of the proposed
model with respect to its hyperparameters, we conducted a series of
experiments. We investigated the effect of using different numbers of ASG modules in the second stage of our model, different values of $K$ in the KNN algorithm, and choosing different percentages of high-score patches on the performance of our model for all the three datasets.

\begin{figure*}[th!]
    \centering
    \includegraphics[width=\textwidth]{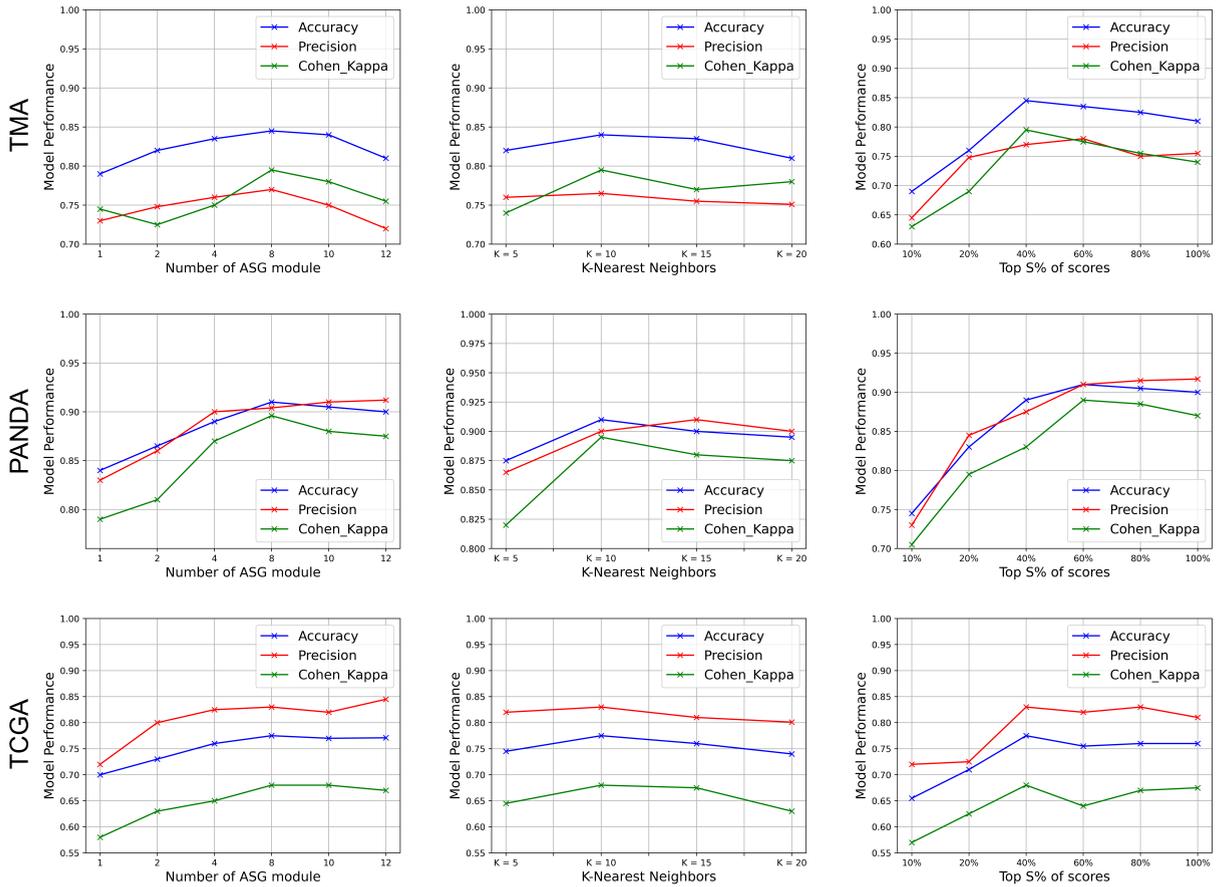}
    \caption{The effect of choosing a different number of ASG modules, different number of neighbors for each node in the graph, and choosing the different percentage of high-score in final metrics for all datasets.}
    \label{fig:ablation}
\end{figure*}

To examine the effect of using different numbers of ASG modules on the performance of the proposed model, we trained the model with up to 12 ASG modules. The performances of the model with different ASG modules on all the datasets are shown in the first column in Figure \ref{fig:ablation}. For the Gleason 2019 challenge  and PANDA datasets, using 8 ASG modules gives us the best performance. However, for TCGA-PRAD dataset, using 12 ASG slightly outperforms the others. The reason can be that the number of instances in the bag and the patch size for TCGA-PRAD is higher than those of the other two datasets and as a result, the constructed graph for TCGA-PRAD is more complex than the others.   But using 12 ASG modules in the model significantly increases the training time and requires more powerful computational resources. Therefore, we stick with using 8 ASG in our experiment and omit the small difference in the accuracy.

We also evaluated the performance of the proposed model using different $K$ values to build connections in the graph and different percentages of high-score patches. The results are shown in the second and third columns in Figure \ref{fig:ablation}, respectively. As the results show, changing the $K$ does not change the accuracy significantly, especially in  Gleason 2019 challenge dataset and TCGA-PRAD, and in all datasets $K$ equal to 10 obtained the best results with a small margin. However, choosing the different percentages of high-score patches can significantly change the performance of the model. For example, in the PANDA dataset, the accuracy increased from 75\% to about 90\% by using 60\% of high-score patches instead of 10\%. On the other hand, using a higher percentage of patches, introduces more noise to the model and it may affect the model's performance.

\section{Conclusions}
In this work, we proposed a novel weakly supervised algorithm to perform Gleason grading of prostate WSIs. Our proposed framework is composed of an attention module to extract and rate the discriminative area in WSI and an augmented self-attention graph neural network (ASG)
model to classify the Gleason grade of WSIs using the graphs constructed on high-score patches. The proposed model was trained and evaluated on three public datasets and cross validated on an independent dataset. The results indicate that using the structured graph on the discriminative areas of WSIs with the patch feature vectors as node features increases the classification accuracy compared with not using the graphs, showing that learning the effect of each patch on other patches in a WSI can be useful in grade classification. Moreover, the results show that the proposed approach outperforms different baseline models in Gleason grade classification.
The promising results presented in this work attest that by using the proposed weakly supervised learning, the gigapixel WSIs can be classified with high accuracy without a need for time-consuming and expensive WSI annotation.


\section{Acknowledgement}
The authors would like to thank Dr. Michael Blechner for providing constructive ideas during the project. This work utilized the Extreme Science and Engineering Discovery Environment (XSEDE), which is supported by National Science Foundation Grant No ACI-1548562. XSEDE resources Stampede 2 and Ranch at the Texas Advanced Computing Center and Bridges at the Pittsburg Supercomputing Center through allocation TG-MCB180008 were used.




\bibliographystyle{cas-model2-names}

\bibliography{cas-refs}

\begin{thebibliography}{48}
\expandafter\ifx\csname natexlab\endcsname\relax\def\natexlab#1{#1}\fi
\providecommand{\url}[1]{\texttt{#1}}
\providecommand{\href}[2]{#2}
\providecommand{\path}[1]{#1}
\providecommand{\DOIprefix}{doi:}
\providecommand{\ArXivprefix}{arXiv:}
\providecommand{\URLprefix}{URL: }
\providecommand{\Pubmedprefix}{pmid:}
\providecommand{\doi}[1]{\href{http://dx.doi.org/#1}{\path{#1}}}
\providecommand{\Pubmed}[1]{\href{pmid:#1}{\path{#1}}}
\providecommand{\bibinfo}[2]{#2}
\ifx\xfnm\relax \def\xfnm[#1]{\unskip,\space#1}\fi
\bibitem[{Allsbrook~Jr et~al.(2001)Allsbrook~Jr, Mangold, Johnson, Lane, Lane
  and Epstein}]{allsbrook2001interobserver}
\bibinfo{author}{Allsbrook~Jr, W.C.}, \bibinfo{author}{Mangold, K.A.},
  \bibinfo{author}{Johnson, M.H.}, \bibinfo{author}{Lane, R.B.},
  \bibinfo{author}{Lane, C.G.}, \bibinfo{author}{Epstein, J.I.},
  \bibinfo{year}{2001}.
\newblock \bibinfo{title}{Interobserver reproducibility of gleason grading of
  prostatic carcinoma: general pathologist}.
\newblock \bibinfo{journal}{Human pathology} \bibinfo{volume}{32},
  \bibinfo{pages}{81--88}.
\bibitem[{Arvaniti et~al.(2018)Arvaniti, Fricker, Moret, Rupp, Hermanns,
  Fankhauser, Wey, Wild, Rueschoff and Claassen}]{arvaniti2018automated}
\bibinfo{author}{Arvaniti, E.}, \bibinfo{author}{Fricker, K.S.},
  \bibinfo{author}{Moret, M.}, \bibinfo{author}{Rupp, N.},
  \bibinfo{author}{Hermanns, T.}, \bibinfo{author}{Fankhauser, C.},
  \bibinfo{author}{Wey, N.}, \bibinfo{author}{Wild, P.J.},
  \bibinfo{author}{Rueschoff, J.H.}, \bibinfo{author}{Claassen, M.},
  \bibinfo{year}{2018}.
\newblock \bibinfo{title}{Automated gleason grading of prostate cancer tissue
  microarrays via deep learning}.
\newblock \bibinfo{journal}{Scientific reports} \bibinfo{volume}{8},
  \bibinfo{pages}{1--11}.
\bibitem[{Bai et~al.(2022a)Bai, Jin, Jin, Wang, Yang and
  Nabavi}]{bai2022applying}
\bibinfo{author}{Bai, J.}, \bibinfo{author}{Jin, A.}, \bibinfo{author}{Jin,
  A.}, \bibinfo{author}{Wang, T.}, \bibinfo{author}{Yang, C.},
  \bibinfo{author}{Nabavi, S.}, \bibinfo{year}{2022}a.
\newblock \bibinfo{title}{Applying graph convolution neural network in digital
  breast tomosynthesis for cancer classification}, in:
  \bibinfo{booktitle}{Proceedings of the 13th ACM International Conference on
  Bioinformatics, Computational Biology and Health Informatics}, pp.
  \bibinfo{pages}{1--10}.
\bibitem[{Bai et~al.(2022b)Bai, Jin, Wang, Yang and Nabavi}]{bai2022feature}
\bibinfo{author}{Bai, J.}, \bibinfo{author}{Jin, A.}, \bibinfo{author}{Wang,
  T.}, \bibinfo{author}{Yang, C.}, \bibinfo{author}{Nabavi, S.},
  \bibinfo{year}{2022}b.
\newblock \bibinfo{title}{Feature fusion siamese network for breast cancer
  detection comparing current and prior mammograms}.
\newblock \bibinfo{journal}{Medical Physics} \bibinfo{volume}{49},
  \bibinfo{pages}{3654--3669}.
\bibitem[{Bai et~al.(2022c)Bai, Li and Nabavi}]{bai2022semi}
\bibinfo{author}{Bai, J.}, \bibinfo{author}{Li, B.}, \bibinfo{author}{Nabavi,
  S.}, \bibinfo{year}{2022}c.
\newblock \bibinfo{title}{Semi-supervised classification of disease prognosis
  using cr images with clinical data structured graph}, in:
  \bibinfo{booktitle}{Proceedings of the 13th ACM International Conference on
  Bioinformatics, Computational Biology and Health Informatics}, pp.
  \bibinfo{pages}{1--9}.
\bibitem[{Bai et~al.(2021)Bai, Posner, Wang, Yang and Nabavi}]{bai2021applying}
\bibinfo{author}{Bai, J.}, \bibinfo{author}{Posner, R.}, \bibinfo{author}{Wang,
  T.}, \bibinfo{author}{Yang, C.}, \bibinfo{author}{Nabavi, S.},
  \bibinfo{year}{2021}.
\newblock \bibinfo{title}{Applying deep learning in digital breast
  tomosynthesis for automatic breast cancer detection: A review}.
\newblock \bibinfo{journal}{Medical image analysis} \bibinfo{volume}{71},
  \bibinfo{pages}{102049}.
\bibitem[{Behzadi and Ilie{\c{s}}(2021)}]{behzadi2021real}
\bibinfo{author}{Behzadi, M.M.}, \bibinfo{author}{Ilie{\c{s}}, H.T.},
  \bibinfo{year}{2021}.
\newblock \bibinfo{title}{Real-time topology optimization in 3d via deep
  transfer learning}.
\newblock \bibinfo{journal}{Computer-Aided Design} \bibinfo{volume}{135},
  \bibinfo{pages}{103014}.
\bibitem[{Behzadi and Ilie{\c{s}}(2022)}]{behzadi2022gantl}
\bibinfo{author}{Behzadi, M.M.}, \bibinfo{author}{Ilie{\c{s}}, H.T.},
  \bibinfo{year}{2022}.
\newblock \bibinfo{title}{Gantl: Toward practical and real-time topology
  optimization with conditional generative adversarial networks and transfer
  learning}.
\newblock \bibinfo{journal}{Journal of Mechanical Design}
  \bibinfo{volume}{144}.
\bibitem[{Bejnordi et~al.(2017)Bejnordi, Veta, Van~Diest, Van~Ginneken,
  Karssemeijer, Litjens, Van Der~Laak, Hermsen, Manson, Balkenhol
  et~al.}]{bejnordi2017diagnostic}
\bibinfo{author}{Bejnordi, B.E.}, \bibinfo{author}{Veta, M.},
  \bibinfo{author}{Van~Diest, P.J.}, \bibinfo{author}{Van~Ginneken, B.},
  \bibinfo{author}{Karssemeijer, N.}, \bibinfo{author}{Litjens, G.},
  \bibinfo{author}{Van Der~Laak, J.A.}, \bibinfo{author}{Hermsen, M.},
  \bibinfo{author}{Manson, Q.F.}, \bibinfo{author}{Balkenhol, M.}, et~al.,
  \bibinfo{year}{2017}.
\newblock \bibinfo{title}{Diagnostic assessment of deep learning algorithms for
  detection of lymph node metastases in women with breast cancer}.
\newblock \bibinfo{journal}{Jama} \bibinfo{volume}{318},
  \bibinfo{pages}{2199--2210}.
\bibitem[{Bulten et~al.(2020)Bulten, Litjens, Pinckaers, Str{\"o}m, Eklund,
  Kartasalo, Demkin and Dane}]{bulten2020panda}
\bibinfo{author}{Bulten, W.}, \bibinfo{author}{Litjens, G.},
  \bibinfo{author}{Pinckaers, H.}, \bibinfo{author}{Str{\"o}m, P.},
  \bibinfo{author}{Eklund, M.}, \bibinfo{author}{Kartasalo, K.},
  \bibinfo{author}{Demkin, M.}, \bibinfo{author}{Dane, S.},
  \bibinfo{year}{2020}.
\newblock \bibinfo{title}{The panda challenge: Prostate cancer grade assessment
  using the gleason grading system}.
\newblock \bibinfo{journal}{MICCAI challenge} .
\bibitem[{Campanella et~al.(2019)Campanella, Hanna, Geneslaw, Miraflor, Werneck
  Krauss~Silva, Busam, Brogi, Reuter, Klimstra and
  Fuchs}]{campanella2019clinical}
\bibinfo{author}{Campanella, G.}, \bibinfo{author}{Hanna, M.G.},
  \bibinfo{author}{Geneslaw, L.}, \bibinfo{author}{Miraflor, A.},
  \bibinfo{author}{Werneck Krauss~Silva, V.}, \bibinfo{author}{Busam, K.J.},
  \bibinfo{author}{Brogi, E.}, \bibinfo{author}{Reuter, V.E.},
  \bibinfo{author}{Klimstra, D.S.}, \bibinfo{author}{Fuchs, T.J.},
  \bibinfo{year}{2019}.
\newblock \bibinfo{title}{Clinical-grade computational pathology using weakly
  supervised deep learning on whole slide images}.
\newblock \bibinfo{journal}{Nature medicine} \bibinfo{volume}{25},
  \bibinfo{pages}{1301--1309}.
\bibitem[{Cangea et~al.(2018)Cangea, Veli{\v{c}}kovi{\'c}, Jovanovi{\'c}, Kipf
  and Li{\`o}}]{cangea2018towards}
\bibinfo{author}{Cangea, C.}, \bibinfo{author}{Veli{\v{c}}kovi{\'c}, P.},
  \bibinfo{author}{Jovanovi{\'c}, N.}, \bibinfo{author}{Kipf, T.},
  \bibinfo{author}{Li{\`o}, P.}, \bibinfo{year}{2018}.
\newblock \bibinfo{title}{Towards sparse hierarchical graph classifiers}.
\newblock \bibinfo{journal}{arXiv preprint arXiv:1811.01287} .
\bibitem[{Epstein et~al.(2016)Epstein, Egevad, Amin, Delahunt, Srigley and
  Humphrey}]{epstein20162014}
\bibinfo{author}{Epstein, J.I.}, \bibinfo{author}{Egevad, L.},
  \bibinfo{author}{Amin, M.B.}, \bibinfo{author}{Delahunt, B.},
  \bibinfo{author}{Srigley, J.R.}, \bibinfo{author}{Humphrey, P.A.},
  \bibinfo{year}{2016}.
\newblock \bibinfo{title}{The 2014 international society of urological
  pathology (isup) consensus conference on gleason grading of prostatic
  carcinoma}.
\newblock \bibinfo{journal}{The American journal of surgical pathology}
  \bibinfo{volume}{40}, \bibinfo{pages}{244--252}.
\bibitem[{Fitzgerald(2001)}]{fitzgerald2001error}
\bibinfo{author}{Fitzgerald, R.}, \bibinfo{year}{2001}.
\newblock \bibinfo{title}{Error in radiology}.
\newblock \bibinfo{journal}{Clinical radiology} \bibinfo{volume}{56},
  \bibinfo{pages}{938--946}.
\bibitem[{Gleason(1992)}]{gleason1992histologic}
\bibinfo{author}{Gleason, D.F.}, \bibinfo{year}{1992}.
\newblock \bibinfo{title}{Histologic grading of prostate cancer: a
  perspective}.
\newblock \bibinfo{journal}{Human pathology} \bibinfo{volume}{23},
  \bibinfo{pages}{273--279}.
\bibitem[{G{\'o}rriz~Blanch(2017)}]{gorriz2017active}
\bibinfo{author}{G{\'o}rriz~Blanch, M.}, \bibinfo{year}{2017}.
\newblock \bibinfo{title}{Active deep learning for medical imaging
  segmentation}.
\newblock \bibinfo{type}{{B.S.} thesis}. Universitat Polit{\`e}cnica de
  Catalunya.
\bibitem[{Gour et~al.(2022)Gour, Jain and Shankar}]{gour2022application}
\bibinfo{author}{Gour, M.}, \bibinfo{author}{Jain, S.},
  \bibinfo{author}{Shankar, U.}, \bibinfo{year}{2022}.
\newblock \bibinfo{title}{Application of deep learning techniques for prostate
  cancer grading using histopathological images}, in:
  \bibinfo{booktitle}{International Conference on Computer Vision and Image
  Processing}, \bibinfo{organization}{Springer}. pp. \bibinfo{pages}{83--94}.
\bibitem[{He et~al.(2016)He, Zhang, Ren and Sun}]{he2016deep}
\bibinfo{author}{He, K.}, \bibinfo{author}{Zhang, X.}, \bibinfo{author}{Ren,
  S.}, \bibinfo{author}{Sun, J.}, \bibinfo{year}{2016}.
\newblock \bibinfo{title}{Deep residual learning for image recognition}, in:
  \bibinfo{booktitle}{Proceedings of the IEEE conference on computer vision and
  pattern recognition}, pp. \bibinfo{pages}{770--778}.
\bibitem[{Howard et~al.(2017)Howard, Zhu, Chen, Kalenichenko, Wang, Weyand,
  Andreetto and Adam}]{howard2017mobilenets}
\bibinfo{author}{Howard, A.G.}, \bibinfo{author}{Zhu, M.},
  \bibinfo{author}{Chen, B.}, \bibinfo{author}{Kalenichenko, D.},
  \bibinfo{author}{Wang, W.}, \bibinfo{author}{Weyand, T.},
  \bibinfo{author}{Andreetto, M.}, \bibinfo{author}{Adam, H.},
  \bibinfo{year}{2017}.
\newblock \bibinfo{title}{Mobilenets: Efficient convolutional neural networks
  for mobile vision applications}.
\newblock \bibinfo{journal}{arXiv preprint arXiv:1704.04861} .
\bibitem[{Ilse et~al.(2018)Ilse, Tomczak and Welling}]{ilse2018attention}
\bibinfo{author}{Ilse, M.}, \bibinfo{author}{Tomczak, J.},
  \bibinfo{author}{Welling, M.}, \bibinfo{year}{2018}.
\newblock \bibinfo{title}{Attention-based deep multiple instance learning}, in:
  \bibinfo{booktitle}{International conference on machine learning},
  \bibinfo{organization}{PMLR}. pp. \bibinfo{pages}{2127--2136}.
\bibitem[{Jafari-Khouzani and Soltanian-Zadeh(2003)}]{jafari2003multiwavelet}
\bibinfo{author}{Jafari-Khouzani, K.}, \bibinfo{author}{Soltanian-Zadeh, H.},
  \bibinfo{year}{2003}.
\newblock \bibinfo{title}{Multiwavelet grading of pathological images of
  prostate}.
\newblock \bibinfo{journal}{IEEE Transactions on Biomedical Engineering}
  \bibinfo{volume}{50}, \bibinfo{pages}{697--704}.
\bibitem[{Karimi et~al.(2019)Karimi, Nir, Fazli, Black, Goldenberg and
  Salcudean}]{karimi2019deep}
\bibinfo{author}{Karimi, D.}, \bibinfo{author}{Nir, G.},
  \bibinfo{author}{Fazli, L.}, \bibinfo{author}{Black, P.C.},
  \bibinfo{author}{Goldenberg, L.}, \bibinfo{author}{Salcudean, S.E.},
  \bibinfo{year}{2019}.
\newblock \bibinfo{title}{Deep learning-based gleason grading of prostate
  cancer from histopathology images—role of multiscale decision aggregation
  and data augmentation}.
\newblock \bibinfo{journal}{IEEE journal of biomedical and health informatics}
  \bibinfo{volume}{24}, \bibinfo{pages}{1413--1426}.
\bibitem[{Kipf and Welling(2016)}]{kipf2016semi}
\bibinfo{author}{Kipf, T.N.}, \bibinfo{author}{Welling, M.},
  \bibinfo{year}{2016}.
\newblock \bibinfo{title}{Semi-supervised classification with graph
  convolutional networks}.
\newblock \bibinfo{journal}{arXiv preprint arXiv:1609.02907} .
\bibitem[{Kunkel et~al.(2021)Kunkel, Madani, White, Verardi and
  Tarakanova}]{kunkel2021modeling}
\bibinfo{author}{Kunkel, G.}, \bibinfo{author}{Madani, M.},
  \bibinfo{author}{White, S.J.}, \bibinfo{author}{Verardi, P.H.},
  \bibinfo{author}{Tarakanova, A.}, \bibinfo{year}{2021}.
\newblock \bibinfo{title}{Modeling coronavirus spike protein dynamics:
  implications for immunogenicity and immune escape}.
\newblock \bibinfo{journal}{Biophysical Journal} \bibinfo{volume}{120},
  \bibinfo{pages}{5592--5618}.
\bibitem[{Latour et~al.(2008)Latour, Amin, Billis, Egevad, Grignon, Humphrey,
  Reuter, Sakr, Srigley, Wheeler et~al.}]{latour2008grading}
\bibinfo{author}{Latour, M.}, \bibinfo{author}{Amin, M.B.},
  \bibinfo{author}{Billis, A.}, \bibinfo{author}{Egevad, L.},
  \bibinfo{author}{Grignon, D.J.}, \bibinfo{author}{Humphrey, P.A.},
  \bibinfo{author}{Reuter, V.E.}, \bibinfo{author}{Sakr, W.A.},
  \bibinfo{author}{Srigley, J.R.}, \bibinfo{author}{Wheeler, T.M.}, et~al.,
  \bibinfo{year}{2008}.
\newblock \bibinfo{title}{Grading of invasive cribriform carcinoma on prostate
  needle biopsy: an interobserver study among experts in genitourinary
  pathology}.
\newblock \bibinfo{journal}{The American journal of surgical pathology}
  \bibinfo{volume}{32}, \bibinfo{pages}{1532--1539}.
\bibitem[{Lee et~al.(2019)Lee, Lee and Kang}]{lee2019self}
\bibinfo{author}{Lee, J.}, \bibinfo{author}{Lee, I.}, \bibinfo{author}{Kang,
  J.}, \bibinfo{year}{2019}.
\newblock \bibinfo{title}{Self-attention graph pooling}, in:
  \bibinfo{booktitle}{International conference on machine learning},
  \bibinfo{organization}{PMLR}. pp. \bibinfo{pages}{3734--3743}.
\bibitem[{Li et~al.(2018)Li, Li, Sarma, Ho, Shen, Knudsen, Gertych and
  Arnold}]{li2018path}
\bibinfo{author}{Li, W.}, \bibinfo{author}{Li, J.}, \bibinfo{author}{Sarma,
  K.V.}, \bibinfo{author}{Ho, K.C.}, \bibinfo{author}{Shen, S.},
  \bibinfo{author}{Knudsen, B.S.}, \bibinfo{author}{Gertych, A.},
  \bibinfo{author}{Arnold, C.W.}, \bibinfo{year}{2018}.
\newblock \bibinfo{title}{Path r-cnn for prostate cancer diagnosis and gleason
  grading of histological images}.
\newblock \bibinfo{journal}{IEEE transactions on medical imaging}
  \bibinfo{volume}{38}, \bibinfo{pages}{945--954}.
\bibitem[{Liu et~al.(2022)Liu, Yang, Mohammadi, Song, Bi and
  Wang}]{liu2022machine}
\bibinfo{author}{Liu, Q.}, \bibinfo{author}{Yang, M.},
  \bibinfo{author}{Mohammadi, K.}, \bibinfo{author}{Song, D.},
  \bibinfo{author}{Bi, J.}, \bibinfo{author}{Wang, G.}, \bibinfo{year}{2022}.
\newblock \bibinfo{title}{Machine learning crop yield models based on
  meteorological features and comparison with a process-based model}.
\newblock \bibinfo{journal}{Artificial Intelligence for the Earth Systems}
  \bibinfo{volume}{1}, \bibinfo{pages}{e220002}.
\bibitem[{Lu et~al.(2022)Lu, Chen, Kong, Lipkova, Singh, Williamson, Chen and
  Mahmood}]{lu2022federated}
\bibinfo{author}{Lu, M.Y.}, \bibinfo{author}{Chen, R.J.},
  \bibinfo{author}{Kong, D.}, \bibinfo{author}{Lipkova, J.},
  \bibinfo{author}{Singh, R.}, \bibinfo{author}{Williamson, D.F.},
  \bibinfo{author}{Chen, T.Y.}, \bibinfo{author}{Mahmood, F.},
  \bibinfo{year}{2022}.
\newblock \bibinfo{title}{Federated learning for computational pathology on
  gigapixel whole slide images}.
\newblock \bibinfo{journal}{Medical image analysis} \bibinfo{volume}{76},
  \bibinfo{pages}{102298}.
\bibitem[{Lu et~al.(2021)Lu, Williamson, Chen, Chen, Barbieri and
  Mahmood}]{lu2021data}
\bibinfo{author}{Lu, M.Y.}, \bibinfo{author}{Williamson, D.F.},
  \bibinfo{author}{Chen, T.Y.}, \bibinfo{author}{Chen, R.J.},
  \bibinfo{author}{Barbieri, M.}, \bibinfo{author}{Mahmood, F.},
  \bibinfo{year}{2021}.
\newblock \bibinfo{title}{Data-efficient and weakly supervised computational
  pathology on whole-slide images}.
\newblock \bibinfo{journal}{Nature biomedical engineering} \bibinfo{volume}{5},
  \bibinfo{pages}{555--570}.
\bibitem[{Lucas et~al.(2019)Lucas, Jansen, Savci-Heijink, Meijer, de~Boer, van
  Leeuwen, de~Bruin and Marquering}]{lucas2019deep}
\bibinfo{author}{Lucas, M.}, \bibinfo{author}{Jansen, I.},
  \bibinfo{author}{Savci-Heijink, C.D.}, \bibinfo{author}{Meijer, S.L.},
  \bibinfo{author}{de~Boer, O.J.}, \bibinfo{author}{van Leeuwen, T.G.},
  \bibinfo{author}{de~Bruin, D.M.}, \bibinfo{author}{Marquering, H.A.},
  \bibinfo{year}{2019}.
\newblock \bibinfo{title}{Deep learning for automatic gleason pattern
  classification for grade group determination of prostate biopsies}.
\newblock \bibinfo{journal}{Virchows Archiv} \bibinfo{volume}{475},
  \bibinfo{pages}{77--83}.
\bibitem[{Madani et~al.(2022a)Madani, Behzadi and Nabavi}]{madani2022role}
\bibinfo{author}{Madani, M.}, \bibinfo{author}{Behzadi, M.M.},
  \bibinfo{author}{Nabavi, S.}, \bibinfo{year}{2022}a.
\newblock \bibinfo{title}{The role of deep learning in advancing breast cancer
  detection using different imaging modalities: A systematic review}.
\newblock \bibinfo{journal}{Cancers} \bibinfo{volume}{14},
  \bibinfo{pages}{5334}.
\bibitem[{Madani et~al.(2022b)Madani, Behzadi, Song, Ilies and
  Tarakanova}]{madani2022improved}
\bibinfo{author}{Madani, M.}, \bibinfo{author}{Behzadi, M.M.},
  \bibinfo{author}{Song, D.}, \bibinfo{author}{Ilies, H.T.},
  \bibinfo{author}{Tarakanova, A.}, \bibinfo{year}{2022}b.
\newblock \bibinfo{title}{Improved inter-residue contact prediction via a
  hybrid generative model and dynamic loss function}.
\newblock \bibinfo{journal}{Computational and Structural Biotechnology Journal}
  \bibinfo{volume}{20}, \bibinfo{pages}{6138--6148}.
\bibitem[{Madani et~al.(2021)Madani, Lin and Tarakanova}]{madani2021dsressol}
\bibinfo{author}{Madani, M.}, \bibinfo{author}{Lin, K.},
  \bibinfo{author}{Tarakanova, A.}, \bibinfo{year}{2021}.
\newblock \bibinfo{title}{Dsressol: A sequence-based solubility predictor
  created with dilated squeeze excitation residual networks}.
\newblock \bibinfo{journal}{International Journal of Molecular Sciences}
  \bibinfo{volume}{22}, \bibinfo{pages}{13555}.
\bibitem[{Nguyen et~al.(2012)Nguyen, Sabata and Jain}]{nguyen2012prostate}
\bibinfo{author}{Nguyen, K.}, \bibinfo{author}{Sabata, B.},
  \bibinfo{author}{Jain, A.K.}, \bibinfo{year}{2012}.
\newblock \bibinfo{title}{Prostate cancer grading: Gland segmentation and
  structural features}.
\newblock \bibinfo{journal}{Pattern Recognition Letters} \bibinfo{volume}{33},
  \bibinfo{pages}{951--961}.
\bibitem[{Nir et~al.(2018)Nir, Hor, Karimi, Fazli, Skinnider, Tavassoli,
  Turbin, Villamil, Wang, Wilson et~al.}]{nir2018automatic}
\bibinfo{author}{Nir, G.}, \bibinfo{author}{Hor, S.}, \bibinfo{author}{Karimi,
  D.}, \bibinfo{author}{Fazli, L.}, \bibinfo{author}{Skinnider, B.F.},
  \bibinfo{author}{Tavassoli, P.}, \bibinfo{author}{Turbin, D.},
  \bibinfo{author}{Villamil, C.F.}, \bibinfo{author}{Wang, G.},
  \bibinfo{author}{Wilson, R.S.}, et~al., \bibinfo{year}{2018}.
\newblock \bibinfo{title}{Automatic grading of prostate cancer in digitized
  histopathology images: Learning from multiple experts}.
\newblock \bibinfo{journal}{Medical image analysis} \bibinfo{volume}{50},
  \bibinfo{pages}{167--180}.
\bibitem[{Ot{\'a}lora et~al.(2021)Ot{\'a}lora, Marini, M{\"u}ller and
  Atzori}]{otalora2021combining}
\bibinfo{author}{Ot{\'a}lora, S.}, \bibinfo{author}{Marini, N.},
  \bibinfo{author}{M{\"u}ller, H.}, \bibinfo{author}{Atzori, M.},
  \bibinfo{year}{2021}.
\newblock \bibinfo{title}{Combining weakly and strongly supervised learning
  improves strong supervision in gleason pattern classification}.
\newblock \bibinfo{journal}{BMC Medical Imaging} \bibinfo{volume}{21},
  \bibinfo{pages}{1--14}.
\bibitem[{Ronneberger et~al.(2015)Ronneberger, Fischer and
  Brox}]{ronneberger2015u}
\bibinfo{author}{Ronneberger, O.}, \bibinfo{author}{Fischer, P.},
  \bibinfo{author}{Brox, T.}, \bibinfo{year}{2015}.
\newblock \bibinfo{title}{U-net: Convolutional networks for biomedical image
  segmentation}, in: \bibinfo{booktitle}{International Conference on Medical
  image computing and computer-assisted intervention},
  \bibinfo{organization}{Springer}. pp. \bibinfo{pages}{234--241}.
\bibitem[{Siegel et~al.(2021)Siegel, Miller, Fuchs, Jemal
  et~al.}]{siegel2021cancer}
\bibinfo{author}{Siegel, R.L.}, \bibinfo{author}{Miller, K.D.},
  \bibinfo{author}{Fuchs, H.E.}, \bibinfo{author}{Jemal, A.}, et~al.,
  \bibinfo{year}{2021}.
\newblock \bibinfo{title}{Cancer statistics, 2021}.
\newblock \bibinfo{journal}{Ca Cancer J Clin} \bibinfo{volume}{71},
  \bibinfo{pages}{7--33}.
\bibitem[{Silva-Rodr{\'\i}guez et~al.(2021)Silva-Rodr{\'\i}guez, Colomer, Dolz
  and Naranjo}]{silva2021self}
\bibinfo{author}{Silva-Rodr{\'\i}guez, J.}, \bibinfo{author}{Colomer, A.},
  \bibinfo{author}{Dolz, J.}, \bibinfo{author}{Naranjo, V.},
  \bibinfo{year}{2021}.
\newblock \bibinfo{title}{Self-learning for weakly supervised gleason grading
  of local patterns}.
\newblock \bibinfo{journal}{IEEE journal of biomedical and health informatics}
  \bibinfo{volume}{25}, \bibinfo{pages}{3094--3104}.
\bibitem[{Singhal et~al.(2022)Singhal, Soni, Bonthu, Chattopadhyay, Samanta,
  Joshi, Jojera, Chharchhodawala, Agarwal, Desai et~al.}]{singhal2022deep}
\bibinfo{author}{Singhal, N.}, \bibinfo{author}{Soni, S.},
  \bibinfo{author}{Bonthu, S.}, \bibinfo{author}{Chattopadhyay, N.},
  \bibinfo{author}{Samanta, P.}, \bibinfo{author}{Joshi, U.},
  \bibinfo{author}{Jojera, A.}, \bibinfo{author}{Chharchhodawala, T.},
  \bibinfo{author}{Agarwal, A.}, \bibinfo{author}{Desai, M.}, et~al.,
  \bibinfo{year}{2022}.
\newblock \bibinfo{title}{A deep learning system for prostate cancer diagnosis
  and grading in whole slide images of core needle biopsies}.
\newblock \bibinfo{journal}{Scientific reports} \bibinfo{volume}{12},
  \bibinfo{pages}{1--11}.
\bibitem[{Smith et~al.(1999)Smith, Zajicek, Werman, Pizov and
  Sherman}]{smith1999similarity}
\bibinfo{author}{Smith, Y.}, \bibinfo{author}{Zajicek, G.},
  \bibinfo{author}{Werman, M.}, \bibinfo{author}{Pizov, G.},
  \bibinfo{author}{Sherman, Y.}, \bibinfo{year}{1999}.
\newblock \bibinfo{title}{Similarity measurement method for the classification
  of architecturally differentiated images}.
\newblock \bibinfo{journal}{Computers and Biomedical Research}
  \bibinfo{volume}{32}, \bibinfo{pages}{1--12}.
\bibitem[{Szegedy et~al.(2016)Szegedy, Vanhoucke, Ioffe, Shlens and
  Wojna}]{szegedy2016rethinking}
\bibinfo{author}{Szegedy, C.}, \bibinfo{author}{Vanhoucke, V.},
  \bibinfo{author}{Ioffe, S.}, \bibinfo{author}{Shlens, J.},
  \bibinfo{author}{Wojna, Z.}, \bibinfo{year}{2016}.
\newblock \bibinfo{title}{Rethinking the inception architecture for computer
  vision}, in: \bibinfo{booktitle}{Proceedings of the IEEE conference on
  computer vision and pattern recognition}, pp. \bibinfo{pages}{2818--2826}.
\bibitem[{Tabesh et~al.(2007)Tabesh, Teverovskiy, Pang, Kumar, Verbel,
  Kotsianti and Saidi}]{tabesh2007multifeature}
\bibinfo{author}{Tabesh, A.}, \bibinfo{author}{Teverovskiy, M.},
  \bibinfo{author}{Pang, H.Y.}, \bibinfo{author}{Kumar, V.P.},
  \bibinfo{author}{Verbel, D.}, \bibinfo{author}{Kotsianti, A.},
  \bibinfo{author}{Saidi, O.}, \bibinfo{year}{2007}.
\newblock \bibinfo{title}{Multifeature prostate cancer diagnosis and gleason
  grading of histological images}.
\newblock \bibinfo{journal}{IEEE transactions on medical imaging}
  \bibinfo{volume}{26}, \bibinfo{pages}{1366--1378}.
\bibitem[{Tan and Le(2019)}]{tan2019efficientnet}
\bibinfo{author}{Tan, M.}, \bibinfo{author}{Le, Q.}, \bibinfo{year}{2019}.
\newblock \bibinfo{title}{Efficientnet: Rethinking model scaling for
  convolutional neural networks}, in: \bibinfo{booktitle}{International
  conference on machine learning}, \bibinfo{organization}{PMLR}. pp.
  \bibinfo{pages}{6105--6114}.
\bibitem[{Vaswani et~al.(2017)Vaswani, Shazeer, Parmar, Uszkoreit, Jones,
  Gomez, Kaiser and Polosukhin}]{vaswani2017attention}
\bibinfo{author}{Vaswani, A.}, \bibinfo{author}{Shazeer, N.},
  \bibinfo{author}{Parmar, N.}, \bibinfo{author}{Uszkoreit, J.},
  \bibinfo{author}{Jones, L.}, \bibinfo{author}{Gomez, A.N.},
  \bibinfo{author}{Kaiser, {\L}.}, \bibinfo{author}{Polosukhin, I.},
  \bibinfo{year}{2017}.
\newblock \bibinfo{title}{Attention is all you need}.
\newblock \bibinfo{journal}{Advances in neural information processing systems}
  \bibinfo{volume}{30}.
\bibitem[{Yanase and Triantaphyllou(2019)}]{yanase2019seven}
\bibinfo{author}{Yanase, J.}, \bibinfo{author}{Triantaphyllou, E.},
  \bibinfo{year}{2019}.
\newblock \bibinfo{title}{The seven key challenges for the future of
  computer-aided diagnosis in medicine}.
\newblock \bibinfo{journal}{International journal of medical informatics}
  \bibinfo{volume}{129}, \bibinfo{pages}{413--422}.
\bibitem[{Zhong et~al.(2017)Zhong, Guo, Rechsteiner, R{\"u}schoff, Rupp,
  Fankhauser, Saba, Mortezavi, Poyet, Hermanns et~al.}]{zhong2017curated}
\bibinfo{author}{Zhong, Q.}, \bibinfo{author}{Guo, T.},
  \bibinfo{author}{Rechsteiner, M.}, \bibinfo{author}{R{\"u}schoff, J.H.},
  \bibinfo{author}{Rupp, N.}, \bibinfo{author}{Fankhauser, C.},
  \bibinfo{author}{Saba, K.}, \bibinfo{author}{Mortezavi, A.},
  \bibinfo{author}{Poyet, C.}, \bibinfo{author}{Hermanns, T.}, et~al.,
  \bibinfo{year}{2017}.
\newblock \bibinfo{title}{A curated collection of tissue microarray images and
  clinical outcome data of prostate cancer patients}.
\newblock \bibinfo{journal}{Scientific data} \bibinfo{volume}{4},
  \bibinfo{pages}{1--9}.

\end{thebibliography}





\end{document}